# Dendrite fragmentation by catastrophic elastic remelting

**S. Ananiev, P. Nikrityuk, K. Eckert**

Dresden University of Technology

Institute for Aerospace Engineering (IAE)

Marschnerstraße 32, Zi. 411

D-01062 Dresden, Germany

Tel.: +49-(0)351-463-38099

Fax: +49-(0)351-463-38087

Email: sergey.ananiev@tu-dresden.de

## ABSTRACT

The paper proposes a new fragmentation mechanism of dendrite arms. The theoretical basis of this mechanism is a shift in the thermodynamical equilibrium at the solid-liquid interface due to the presence of elastic energy. This effect is modelled by the generalized Gibbs-Thomson condition [1], where each term is calculated analytically using a simple Bernoulli-Euler beam model. The resulting nonlinear system of ordinary differential equations is integrated in time using a fully implicit scheme. It is demonstrated that there is a critical level of loading, exceeding which causes a catastrophic reduction of the neck cross section leading to dendrite detachment.

## USED SYMBOLS

| | |
|---|---|
| $[\bullet]_L^S$ | – jump of some quantity between phases $\left([\bullet]_L^S \equiv \bullet_S - \bullet_L\right)$, |
| $\nabla$ | – gradient vector $\left(\nabla \equiv \partial_x \vec{e}_x + \partial_y \vec{e}_y + \partial_z \vec{e}_z\right)$, |
| $\cdot$ | – single contraction / scalar product $\left(\vec{u} \cdot \vec{v} \equiv \sum_i u_i v_i\right)$, |
| $:$ | – double contraction $\left(\boldsymbol{\sigma} : \boldsymbol{\varepsilon} \equiv \sum_{ij} \sigma_{ij} \varepsilon_{ij}\right)$, |
| $a, b, c$ | – parameters, |
| $C_\infty$ | – remote concentration, |
| $C_L^*, C_S^*$ | – equilibrium concentration of solute in liquid and solid, |
| $D_L$ | – diffusion coefficient in liquid, |
| $E$ | – Young modulus, |
| $F$ | – analytical function describing the geometry of dendrite arm, |
| $G$ | – total Gibbs free energy density, |
| $G_R$ | – surface part of Gibbs free energy density, |
| $G_{El}$ | – elastic energy as a part of Gibbs free energy density, |
| $G_{Sf}$ | – entropy part of Gibbs free energy density, |
| $G_N$ | – mechanical work as a part of Gibbs free energy density, |
| $I_x$ | – moment of inertia of cross section along x-axis, |
| $K$ | – local curvature $(K = 1/R)$, |
| $L$ | – length of dendrite arm, |
| $M_x$ | – x-component of bending moment vector, |
| $m_L, m_S$ | – slope of liquidus/solidus lines, |
| $\vec{n}$ | – normal vector to some surface, |
| $p_0$ | – atmospheric pressure, |
| $Pe$ | – Péclet number $(Pe \equiv VR/2D_L)$, |
| $R$ | – local curvature radius, |
| $R_{El}$ | – curvature radius due to elastic deformations, |
| $T_M^{(0)}$ | – melting temperature of pure substance in the case of planar interface, |
| $T_M$ | – melting temperature of pure substance in the case of curved interface, |
| $t$ | – time, |
| $V$ | – velocity of S-L front propagation, |
| $\Delta s_f$ | – entropy of fusion per unit volume $\left(\Delta s_f \equiv [s_f]_\beta^\alpha\right)$, |
| $\Delta T$ | – undercooling $\left(\Delta T \equiv T_M - T_M^{(0)}\right)$, |
| $\delta \bullet$ | – infinitesimal variation of some quantity, |
| $\delta v$ | – elementary volume due to variation of interface position, |
| $\delta A$ | – elementary area due to variation of interface position, |
| $\boldsymbol{\varepsilon}$ | – symmetric part of deformation rate tensor $\left(\boldsymbol{\varepsilon} \equiv 1/2\left(\partial\vec{\mathbf{v}}/\partial\vec{\mathbf{x}} + \partial\vec{\mathbf{v}}/\partial\vec{\mathbf{x}}^T\right)\right)$, |
| $[\boldsymbol{\varepsilon}]_L^S \cdot \vec{\mathbf{n}}$ | – projection of the jump of strain tensor, |
| $\Omega_C$ | – solute supersaturation $\left(\Omega_C \equiv \left(C_L^* - C_0\right)/\left(C_L^* - C_S^*\right)\right)$, |
| $\gamma_{SL}$ | – surface energy density between phases, |
| $\eta_{El}, \eta_T, \eta_{Cr}$ | – proportionality factors, |
| $\nu$ | – Poisson modulus, |
| $\boldsymbol{\sigma}$ | – Cauchy stress tensor, |
| $\boldsymbol{\sigma}_L \cdot \vec{\mathbf{n}}$ | – normal traction (projection of the stress tensor acting in the liquid phase), |
| $\tau_0$ | – elasticity limit of material (also referred as structural/theoretical strength). |

## 1 INTRODUCTION

Alloys with a fine-grain-structure, associated with superior mechanical properties, are the ultimate goal of a solidification process. It is generally accepted that the grain refinement is caused by dendrite fragmentation ([2], [3]). There is however a debate about the mechanism behind this phenomenon. Several mechanisms were proposed and later supported by experimental observations: coarsening, solute enhancement and recalescence ([4], [5], [6]). Since the work of Pilling & Hellawell [7] the *mechanical* fragmentation is not considered to be important for grain refinement. It was estimated by the authors that even for extremely elongated dendrites and typical flow velocities $\left(L \sim 200\mu m,\ R^{root} \sim 5\mu m,\ V \sim 10^{-2} m/sec\right)$ the mechanical stresses acting in the dendrite neck remain well below the elasticity limit of material: $\sigma^{MAX}/\tau_0 \sim 0.1$. To remedy the situation it was proposed in [8] that under rapid solidification conditions the flow velocities can be large enough to cause the inelastic deformations in the dendrite neck. The authors however did not take into account that the elasticity limit of material at the microscale must be close to the *theoretical* value: $\tau_0 \sim 10^{-1} E$ ($E \sim 100 GPa$ is the Young modulus). They based their analysis on the *structural* strength, which is two orders of magnitude smaller: $\tau_0 \sim 10^{-3} E$. The difference between these two strengths was clarified in the pioneering work of Griffith [9] (see also [10]). He assumed that at the macroscale there are always some surface defects like flaws or cracks, which are responsible for fracture under considerably lower tensile stress, than the theoretical strength. The theory was supported by experiments, where it was demonstrated that the strength of thin glass fibres increased dramatically (up to 20 times) as their diameter decreased from $10^{-3} m$ to $10^{-6} m$. It seems to us that in growing dendrites such surface defects can not exist, because they would immediately remelt due to the severe curvature variations. Therefore it can be expected that the strength of dendrites must be at least 10 times higher, than the values used in [7], [8]. This leaves no hope for success of pure mechanical theory of fragmentation.

In spite of the theoretical difficulties there is a compelling experimental evidence that dendrites are bent mechanically even under normal solidification conditions [11]. This provides a motivation to look for a principally new fragmentation mechanism, which accounts for the presence of mechanical stresses. Our idea is based on the well known fact that the presence of elastic energy density affects the thermodynamical equilibrium at the solid surfaces. As an example of such an interaction an increase in corrosion rates of mechanically stressed steel elements compared to the unstressed ones can be mentioned [12]. We found out that if this effect is taken into account, then there is a critical loading level, exceeding which causes a catastrophic remelting of the dendrite neck cross section. This process occurs within a very short period of time, which makes it to look very similar to the native mechanical fragmentation. The physics behind it is however completely different – it is of *constitutional remelting* type.

To illustrate our idea we proceed as follows. In Section 2 the generalized Gibbs-Thomson (GGT) condition for pure material is introduced. It constitutes a theoretical foundation of our work. In Section 3, a simplified geometrical model of dendrite arm is presented. It allows explicit analytical expressions for the each term in the GGT condition and to derive a system of ordinary differential equations, which is done in Section 4. With its help a time evolution of the dendrite neck cross section, including the possibility of the *catastrophic elastic remelting (CER)*, is completely described. Finally, in Section 5, the critical value of mechanical loading will be found numerically for typical dendrite arm diameter $(R_0 = 5 \mu m)$. It is demonstrated that for loading above this value the behavior of the neck cross section is characterized by the catastrophic remelting rates. The proposed CER mechanism is summarized in Sections 6 and 7.

## 2    GENERALIZED GIBBS-THOMSON CONDITION

In the thermodynamics of solidification the effect of the mechanical stresses is described by the generalized Gibbs-Thomson condition, derived by Larché & Cahn [1] (see also [13]). Following this work, the total variation of free Gibbs free energy due to variation of the phase boundary reads:

$$\delta G = \underbrace{\left(T_M - T_M^{(0)}\right)\left[s_f\right]_L^S \delta v}_{\delta G_{Sf}} + \underbrace{\gamma_{SL} K \delta v}_{\delta G_R} + \frac{1}{2}\underbrace{\left[\boldsymbol{\sigma}:\boldsymbol{\varepsilon}\right]_L^S \delta v}_{\delta G_{El}} - \underbrace{\left(\boldsymbol{\sigma}_L \cdot \vec{\mathbf{n}} + \gamma_{SL} K \vec{\mathbf{n}}\right) \cdot \left[\boldsymbol{\varepsilon}\right]_L^S \cdot \vec{\mathbf{n}} \delta v}_{\delta G_N}. \qquad (1)$$

Under condition of thermodynamical equilibrium this variation must vanish $(\delta G = 0)$, which allows us to write the expression for local undercooling in the convenient form:

$$\Delta T = -\frac{1}{\Delta s_f}\left(\gamma_{SL} K + \frac{1}{2}\left[\boldsymbol{\sigma}:\boldsymbol{\varepsilon}\right]_L^S - \left(\boldsymbol{\sigma}_L \cdot \vec{\mathbf{n}} + \gamma_{SL} K \vec{\mathbf{n}}\right) \cdot \left[\boldsymbol{\varepsilon}\right]_L^S \cdot \vec{\mathbf{n}}\right) \qquad (2)$$

The physical meaning of all terms is illustrated in Figure 1. The first, $\delta G_R$, is the standard Gibbs-Thomson effect representing the additional surface energy, which is created during the variation of a curved phase boundary. The other two terms, $\delta G_{Sf}$ and $\delta G_{El}$, resulting from the phase change in the elementary volume $\delta v$: the liquid phase is replaced by the solid phase with corresponding elastic energy and entropy density. The last term, $\delta G_N$, represents change in the elastic energy density due to the variation of displacement field (mechanical work).

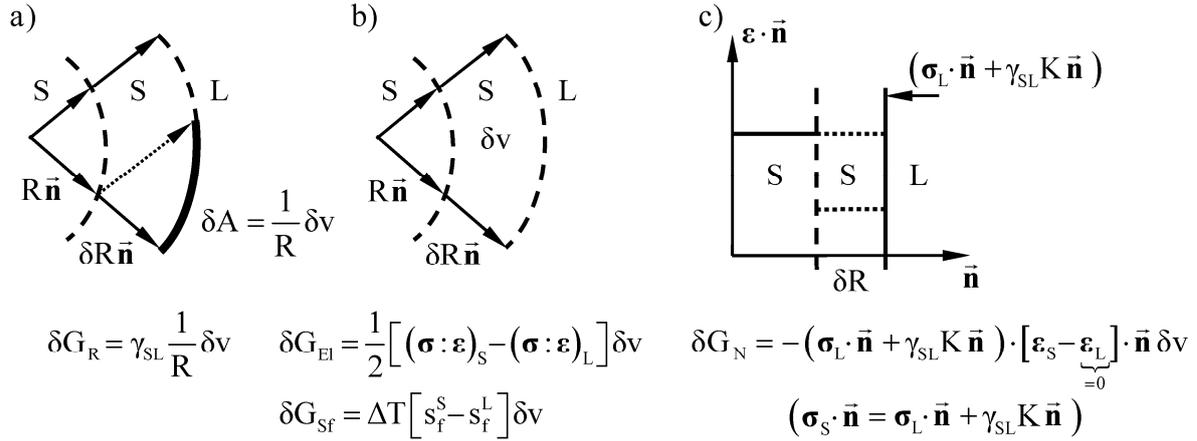

**Figure 1** Illustration of variation of the Gibbs free energy due to variation of the phase boundary: a) additional surface energy; b) change in the elastic energy and in the entropy density in volume $\delta v$; c) change in the elastic energy density due to the variation of displacement field.

## 3   COMPUTATIONAL MODEL OF A DENDRITE ARM

In order to apply Equation (2) to the fragmentation of dendrite arms it is necessary to model all undercooling terms. Following [7], we adopt a simple beam model. This assumption is justified by the typical dimensions of the dendrite arm: $L/R^{root} \sim 10 \div 100$. It can be demonstrated by means of 3D elasticity theory that for such geometries the deformation and stress states are mainly due to the beam bending, where the shear stresses can be neglected. Furthermore, we are interested in the elastic energy density <u>at the surface</u>, see Figure 1, where the shear stresses are equal to zero in the most cases. This leads to the classical Bernoulli-Euler formula [14] connecting the curvature of beam middle line with the bending moment at this point. The physical source for bending moment $M_x$ can be, for example, the buoyancy force.

$$\frac{1}{R_{El}} = \frac{M_x}{EI_x} \tag{3}$$

The further treatment rests on two simplifications. First, we restrict our attention to the neck of dendrite arm. It is obvious that bending moment $M_x$ has a maximal value there while the moment of inertia $I_x$ achieve its minimum. Therefore, if we assume that the mechanical stresses are responsible for detachment of dendrite arms it can happen only at the neck of dendrite. Second, we assume that the neck cross section of the dendrite, at any instant of time, has the shape of an ellipse, see Figure 2, whose semi-axis can move forwards or backwards with velocities depending on the local undercooling. These simplifications allow us to describe the complete problem of dendrite fragmentation using just two degrees of freedom – the actual positions of the ellipse semi-axes –, and illustrate the time-dependent behaviour graphically (Section 5).

### 3.1   Geometry

The chosen model geometry of the dendrite arm is shown in Figure 2. It is completely defined by the three parameters: the time-dependent ellipse semi-axes $A(t)$, $B(t)$ and the constant reference radius $R_0$.

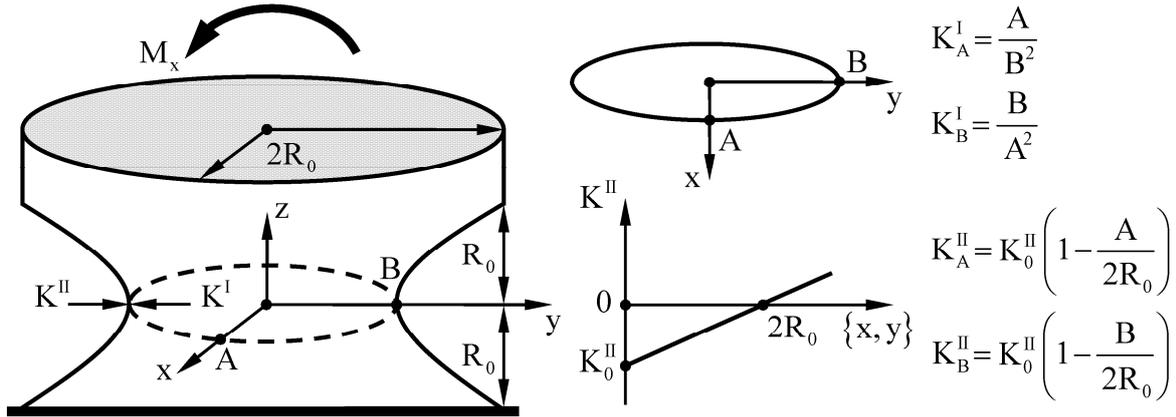

**Figure 2** Geometric model of the neck cross section of the dendrite arm.

The shape of the neck is 3-dimensional with the curvatures of different signs at the semi-axes of ellipse. The first one – $K^I$ is the positive inner curvature of ellipse. The second one – $K^{II}$ is the negative outer curvature of the neck surface. To calculate them analytically the definition of curvature of implicit function is used [15].

$$K(x,y) \equiv \nabla \cdot \vec{n} = \nabla \cdot \frac{\nabla F}{\sqrt{\nabla F \cdot \nabla F}} = \frac{F_{,xx}F_{,y}^2 - 2F_{,xy}F_{,x}F_{,y} + F_{,yy}F_{,x}^2}{\left(F_{,x}^2 + F_{,y}^2\right)^{3/2}} \qquad (4)$$

While the implicit function of the cross section is already known $F^I(x,z) \equiv (x/A)^2 - (y/B)^2 - 1 = 0$, its counterpart for neck surface has still to be defined. It is not necessary to find an analytical form for the whole neck surface. Only the curvature values at the ellipse semi-axes are needed. We assume that in the $x-z$ and $y-z$ planes the neck surface is described by a parabola: $F^{II}(x,z) \equiv x - (az^2 + bz + c) = 0$. Its coefficients can be expressed via primary parameters of neck geometry: $a = 2/R_0 - A/R_0^2$, $b = 0$, $c = A$. The final expression for the outer curvature turned out to be a linear function with respect to the semi-axes of ellipse $\left(K_0^{II} \equiv 4/R_0\right)$:

$$K^I(A,0) \equiv K_A^I = \frac{A}{B^2} \quad K^{II}(A,0) \equiv K_A^{II} = K_0^{II}\left(1 - \frac{A}{2R_0}\right)$$
$$K^I(0,B) \equiv K_B^I = \frac{B}{A^2} \quad K^{II}(0,B) \equiv K_B^{II} = K_0^{II}\left(1 - \frac{B}{2R_0}\right) \qquad (5)$$

Up to this point, the neck cross section had a double symmetry, where the points $(0,B)/(0,-B)$ and $(A,0)/(-A,0)$ were indistinguishable. The situation changes if the change of curvature due to elastic deformations described by (3) is taken into account. At the points $(0,B)/(0,-B)$ it has opposite signs, while its contribution is zero at the $(A,0)/(-A,0)$ points. This is illustrated in Figure 3.

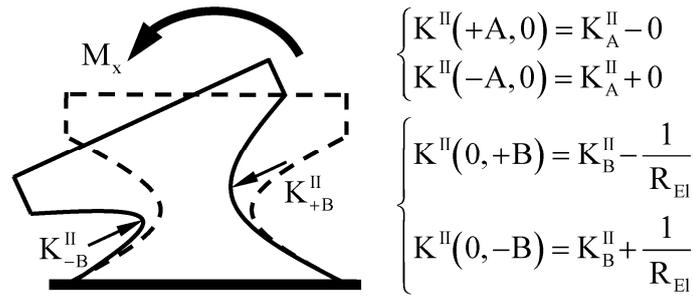

**Figure 3** Effect of elastic deformations on the outer curvature.

The estimation of the magnitude of elastic curvature shows, however, that it can be neglected. Assume the maximal stress to be a given fraction of the Young modulus $\sigma_{zz}^{MAX} \sim \eta_{El} E$. Then, it turns out that the relation between the outer curvature $K^{II}$ and its elastic counterpart is equal to the fraction coefficient $\eta_{El}$.

$$\begin{aligned} \sigma_{zz}^{MAX} \sim \eta_{El} E = \frac{M_x}{I_x} y^{MAX} \\ \frac{M_x}{EI_x} = \frac{1}{R_{El}} \to \frac{1}{R_{El}} = \frac{\eta_{El}}{y^{MAX}} \end{aligned} \to \begin{cases} \frac{1}{R_{El}} \sim \frac{\eta_{El}}{R_0} \\ \frac{1}{R^{II}} \sim \frac{1}{R_0} \end{cases} \to \frac{1/R_{El}}{1/R^{II}} \sim \eta_{El} \qquad (6)$$

Later, in Equation (19), it will be shown that catastrophic elastic remelting takes place at the values $\eta_{El} \sim 10^{-3}$. Therefore the effect of elastic deformations can be neglected. Thus the double symmetry of the neck cross section is preserved and the number of independent variables can be kept by two.

In other cases, where the deformations are large and/or the initial beam geometry does not have any curvature (i.e. it is formed by the straight lines), the effect of elastic deformations cannot be neglected. This could be a natural explanation for the experimentally observed fact that the steel corrosion rates are different at the concave and convex sides of the elastically bent beam. Usually this fact is attributed to the dependence of chemical potential on the <u>sign</u> of elastic strain [16].

## 3.2 Stress state

After having derived analytical expressions for the curvature terms in the GGT condition (2) we now proceed with the elastic ones. According to the chosen Bernoulli-Euler beam model and chosen direction of the applied loading ($M_x$ in Figure 2) the stress state in the neck cross section is a linear function of the y-coordinate, only.

$$\begin{aligned} \sigma_{zz}(x,y) = \frac{M_x}{I_x} y \\ I_x = \frac{\pi}{4} AB^3 \end{aligned} \to \begin{aligned} \sigma_{zz}(A,0) = 0 \\ \sigma_{zz}(0,B) = \frac{4M_x}{\pi AB^2} \end{aligned} \qquad (7)$$

In general, there are also some additional terms in the stress tensor coming from Laplace and atmospheric pressure: $p_0 + \gamma_{SL}(K^I - K^{II})$. The estimation of their magnitude demonstrates, however, that they can be neglected. The surface energy, which is equal to the surface stress in the case of isotropy, is $\gamma_{SL} \sim 0.1 \, \text{N/m}$ for Al-Cu alloys. The difference between both curvatures at the semi-axes of the neck cross section is $(K^I - K^{II}) \sim 10^6 \, \text{m}^{-1}$. The resulting Laplace pressure is therefore $\gamma_{SL}(K^I - K^{II}) \sim 0.1 \, \text{MPa}$. The same order of magnitude also has the atmospheric pressure $p_0$. Comparison of these values with the critical bending stress from (19) $\sigma_{zz} \sim 50 \, \text{MPa}$ demonstrates that they are negligibly small. Thus, during the calculation of the jump in the elastic energy density on the S-L interface only the bending stresses have to be taken into account. The elastic energy in the liquid is obviously zero.

$$\frac{1}{2}[\boldsymbol{\sigma}:\boldsymbol{\varepsilon}]_L^S\bigg|_{(0,B)} = \frac{1}{2}\boldsymbol{\sigma}_S:\boldsymbol{\varepsilon}_S - \underbrace{\frac{1}{2}\boldsymbol{\sigma}_L:\boldsymbol{\varepsilon}_L}_{=0} = \frac{1}{2}\frac{\sigma_{zz}^2}{E} = \frac{8}{E}\left(\frac{M_x}{\pi AB^2}\right)^2 \tag{8}$$

The last term in Equation (2) can be neglected as well. The mechanical work conducted by the Laplace and atmospheric pressure is considerably smaller than the elastic energy density on the S-L interface.

$$[\boldsymbol{\varepsilon}]_L^S\cdot\vec{\mathbf{n}} = \boldsymbol{\varepsilon}_S\cdot\vec{\mathbf{n}} - \underbrace{\boldsymbol{\varepsilon}_L\cdot\vec{\mathbf{n}}}_{=0} = \begin{bmatrix} \nu\sigma_{zz}E^{-1} & 0 & 0 \\ 0 & \nu\sigma_{zz}E^{-1} & 0 \\ 0 & 0 & \sigma_{zz}E^{-1} \end{bmatrix}\cdot\begin{bmatrix}0\\1\\0\end{bmatrix} - \begin{bmatrix}0\\0\\0\end{bmatrix} = \begin{bmatrix}0\\ \nu\sigma_{zz}E^{-1}\\0\end{bmatrix}$$

$$\left(\boldsymbol{\sigma}_L\cdot\vec{\mathbf{n}} + \gamma_{SL}K\vec{\mathbf{n}}\right)\cdot[\boldsymbol{\varepsilon}]_L^S\cdot\vec{\mathbf{n}}\bigg|_{(0,B)} = \left(p_0 + \gamma_{SL}\left(K^I - K^{II}\right)\right)\nu\sigma_{zz}E^{-1} \approx 0, \quad \left(\ll \sigma_{zz}^2 E^{-1}\right) \tag{9}$$

In contrast to the point B, the bending stresses at the point A are equal to zero. Despite this, the effect of Laplace and atmospheric pressure can be neglected here as well. They are considerably smaller than the surface energy. The reason for this is the inverse Young modulus, which is present in both elastic terms in Equation (2). The density of surface energy is $\delta G_R \sim \gamma_{SL}\left(K^I - K^{II}\right) \sim 0.1\,\text{MPa}$. The Laplace and atmospheric pressure are equal to the same value. However, the expression for the elastic energy density is proportional to the square of this value divided by the Young modulus: $\delta G_{El} \sim (0.1\,\text{MPa})^2/70\,\text{GPa} \approx 0.1\,\text{MPa}\cdot 10^{-6}$, which is much smaller than $0.1\,\text{MPa}$. The second elastic term in Equation (2) has the same order of magnitude and can be neglected as well.

$$\frac{1}{2}[\boldsymbol{\sigma}:\boldsymbol{\varepsilon}]_L^S\bigg|_{(A,0)} \sim \frac{\left(p_0 + \gamma_{SL}\left(K^I - K^{II}\right)\right)^2}{E} \approx 0 \qquad \left(\ll \gamma_{SL}\left(K^I - K^{II}\right)\right)$$

$$\left(\boldsymbol{\sigma}_L\cdot\vec{\mathbf{n}} + \gamma_{SL}K\vec{\mathbf{n}}\right)\cdot[\boldsymbol{\varepsilon}]_L^S\cdot\vec{\mathbf{n}}\bigg|_{(A,0)} \sim \nu\frac{\left(p_0 + \gamma_{SL}\left(K^I - K^{II}\right)\right)^2}{E} \approx 0 \quad \left(\ll \gamma_{SL}\left(K^I - K^{II}\right)\right) \tag{10}$$

## 4     DYNAMICS OF THE INTERFACE IN BINARY ALLOYS

Analytical expressions for all terms in the GGT condition (2) were derived in the previous Section. Consequently, the local change in the melting temperature of pure substance is now a nonlinear analytical function of time-dependent ellipse semi-axes: $A(t)$ and $B(t)$.

$$\begin{cases} \Delta T_A(A,B) = -\dfrac{\gamma_{SL}}{\Delta s_f}\left(\dfrac{A}{B^2} - K_0^{II}\left(1 - \dfrac{A}{2R_0}\right)\right) \\[2mm] \Delta T_B(A,B) = -\dfrac{\gamma_{SL}}{\Delta s_f}\left(\dfrac{B}{A^2} - K_0^{II}\left(1 - \dfrac{B}{2R_0}\right) + \dfrac{8}{E\gamma_{SL}}\left(\dfrac{M_x}{\pi AB^2}\right)^2\right) \end{cases} \tag{11}$$

To apply Equation (11) to the solidification of binary alloys, the local undercoolings have to be related to the solute distribution at the S-L interface. It can be shown, that in ideal binary mixtures the small curvature term does not change the slopes of liquidus/solidus lines [17]. Here we assume that the elastic undercooling is small as well, which implies that the new liquidus/solidus lines are obtained by a parallel shift by the amount $\Delta T_{\{A,B\}}$ along the temperature axis (Figure 4a). Hence, the change in the equilibrium solute concentration reads:

$$\left(C_L^* - C_\infty\right)_{\{A,B\}} = -\frac{1}{m_L}\Delta T_{\{A,B\}} . \tag{12}$$

Depending on the sign of local undercooling both situations are possible: partial remelting and (re)solidification, which is illustrated in Figure 4b.

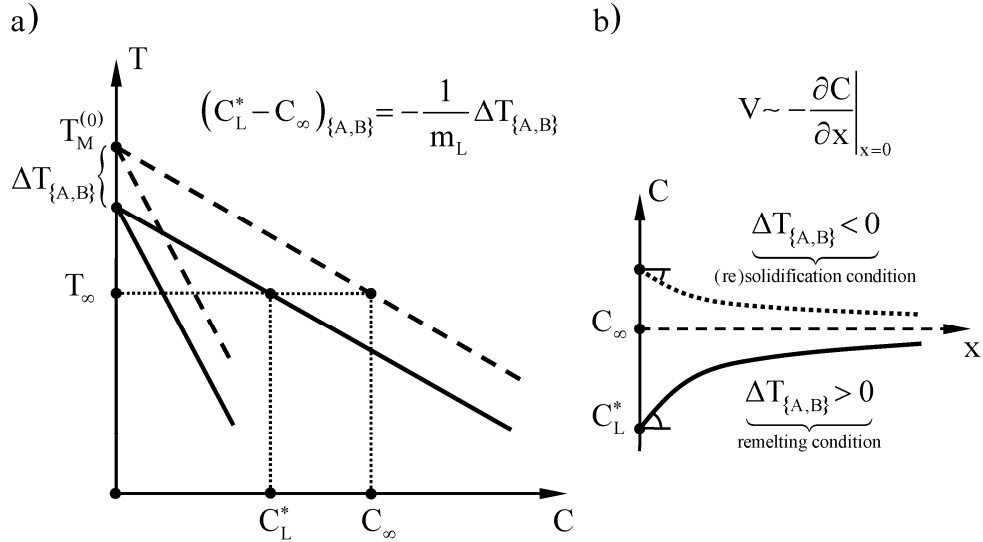

**Figure 4** a) Shift in the phase diagram due to change in the melting temperature; b) distribution of solute in space and corresponding interface velocity depending on the sign of undercooling.

To obtain a time dependent problem it is necessary to relate the velocities of the ellipse semi-axes with the corresponding local undercoolings, i.e. we need the following functional dependencies: $V_A(\Delta T_A)$, $V_B(\Delta T_B)$. The simplest way to define them is to use the hemispherical needle approximation [18], which is based on the steady state analytical solution of the diffusion equation in the case of spherical symmetry. The mass conservation condition in term of the solutal Péclet number and solute supersaturation leads to the nonlinear expression for interface velocity:

$$\frac{VR}{2D_L} = \frac{C_L^* - C_\infty}{C_L^* - C_S^*} \quad \rightarrow \quad V(\Delta T) . \tag{13}$$

The nonlinearity of this Equation comes from the fact that both expressions: $\left(C_L^* - C_\infty\right)$ and $\left(C_L^* - C_S^*\right)$, depend on $\Delta T$. The radius $R$ of hemispherical needle is assumed to be constant and, in our case, to be to equal to $2R_0$. As it was assumed already, the elastic undercooling is small and therefore the Equation (13) can be approximated by its linearization.

$$V(\Delta T) = \frac{D_L}{R_0} \frac{\Delta T}{m_L} \frac{1}{\left(\frac{\Delta T}{m_L} + C_\infty\right)\left(1 - \frac{C_S^*}{C_L^*}\right)} \approx \left.\frac{dV}{d\Delta T}\right|_{\Delta T=0} \Delta T$$

$$\left.\frac{dV}{d\Delta T}\right|_{\Delta T=0} = \frac{D_L}{R_0 m_L C_\infty \left(1 - \frac{C_S^*}{C_L^*}\right)}$$

(14)

## 5  ANALYSIS OF THE FINAL SYSTEM OF EQUATIONS

The final system of equations which completely describes the time evolution of the dendrite neck cross section now reads as follows:

$$\begin{cases} V_A = -\eta_T \Delta T_A(A,B) \\ V_B = -\eta_T \Delta T_B(A,B) \\ \eta_T \equiv \frac{\gamma_{SL}}{\Delta s_f} \left.\frac{dV}{d\Delta T}\right|_{\Delta T=0} \end{cases} \rightarrow \begin{cases} \dfrac{dA}{dt} = -\eta_T \left(\dfrac{A}{B^2} - K_0^{II}\left(1 - \dfrac{A}{2R_0}\right)\right) \\ \dfrac{dB}{dt} = -\eta_T \left(\dfrac{B}{A^2} - K_0^{II}\left(1 - \dfrac{B}{2R_0}\right) + \dfrac{8}{E\gamma_{SL}}\left(\dfrac{M_x}{\pi AB^2}\right)^2\right) \end{cases}$$

(15)

Numerical values of all constants in this equation system were calculated using material data for Al-Cu alloys [18]. The constant $\eta_T$ was calculated according to Equation (14). The only geometric constant – the reference radius of dendrite arm $R_0$ – was set to $5\,\mu m$.

$$\eta_T = \frac{3 \cdot 10^{-9}\,[m^2/\text{sec}] \cdot 0,1\,[N/m]}{5 \cdot 10^{-6}\,[m] \cdot 2,6\,[K/\text{wt\%}] \cdot 0,1\,[\text{wt\%}] \cdot (1-0,14) \cdot 10^6\,[N \cdot m/m^3 \cdot K]} \approx 3 \cdot 10^{-10}\,[m^2/\text{sec}]$$

$$E\gamma_{SL} = 70 \cdot 10^9\,[N/m^2] \cdot 1 \cdot 10^{-1}\,[N/m] \approx 7 \cdot 10^9\,[N^2/m^3]$$

$$K_0^{II} = \frac{4}{5 \cdot 10^{-6}\,[m]} \approx 1 \cdot 10^6\,[1/m]$$

(16)

Before we proceed with the numerical analysis some qualitative remarks are in order. First, for moderate loading $(M_x)$ there must be an equilibrium state, characterized by $\Delta T_A = 0\ \&\ \Delta T_B = 0$, and, hence, a time-independent stable neck cross section $A_{,t} = 0\ \&\ B_{,t} = 0$. The reason for this assumption is the presence of the outer curvature $K^{II}$ in the both expressions for undercoolings. Due to its negative sign it plays a stabilizing role in the dynamics of the neck cross section. If mechanical loading is applied, the semi-axis B will decrease (the cross section will partially remelt). This will cause a further increase of the elastic term due to a decrease of moment of inertia $I_x$. This, however, can be compensated by a <u>decrease</u> of the inner curvature and by an <u>increase</u> of the outer curvature (see Figure 2 and compare with [19]).

Second, if there are no equilibrium states, i.e. the curves $\Delta T_A = 0$ and $\Delta T_B = 0$ do not have common points, the remelting must have a *catastrophic* character: the smaller the semi-axes are, the faster the melting rates will be. This type of solution can be identified if we consider the solution of (15) in the limit of $A \rightarrow 0$, $B \rightarrow 0$. Both velocities are negative and tending to the infinity for this limit case.

$$\begin{cases} \dfrac{dA}{dt} \sim -\dfrac{1}{B} \\ \dfrac{dB}{dt} \sim -\dfrac{1}{A^2 B^4} \end{cases} \quad (17)$$

The system (17) posses no analytical solution in closed form. However, from the analytical solution of simple analogous equation, $A_{,t} = -A^{-1} \rightarrow A(t) = \sqrt{2(t_C - t)}$, it can be inferred that the catastrophic behaviour is of root type with $t_C$ being the time of collapse. Numerical results presented in Figure 7 support our guess.

## 5.1 Critical loading level

The foregoing argumentation presumes the existence of a critical loading, where any further increase will lead to catastrophic remelting. A *posteriori* analysis of Figure 5 indicates that in this case the $\Delta T_A = 0$ and $\Delta T_B = 0$ curves have only one point of intersection (implicit plots were done by GNUPLOT). Mathematically this condition means that at the critical equilibrium point both gradients of undercoolings are linearly dependent. The resulting system of equations (18) has to be solved for four unknowns: both semi-axes, the critical bending moment and the proportionality factor (Lagrange multiplier).

$$\begin{cases} 0 = \Delta T_A \\ 0 = \Delta T_B \\ \dfrac{\partial \Delta T_A}{\partial A} = \eta_{Cr} \dfrac{\partial \Delta T_B}{\partial A} \\ \dfrac{\partial \Delta T_A}{\partial B} = \eta_{Cr} \dfrac{\partial \Delta T_B}{\partial B} \end{cases} \rightarrow \begin{bmatrix} A^{Cr} \\ B^{Cr} \\ M_x^{Cr} \\ \eta_{Cr} \end{bmatrix} \quad (18)$$

For numerical solution of this equation system the full Newton method was used [21], where the choice of initial guess was important to achieve quadratic convergence rate. All partial derivatives needed for the linearization of (18) can be easily calculated from (15) applying the chain rule and are not shown here. The numerical results are summarized in Table 1.

**Table 1** Numerical results of Newton method

| Variables | Values |
| --- | --- |
| $A_0$, $B_0$, $M_0$, $\eta_{Cr}$ [μm, μm, Nμm, _] | 10.000000, 10.000000, 0.015000, 0.000000 |
| $A^{Cr}$, $B^{Cr}$, $M_x^{Cr}$, $\eta_{Cr}$ [μm, μm, Nμm, _] | 8.423333, 7.309236, 0.017042, -1.858209 |

The temporal evolutions of the ellipse semi-axes according to (15) are shown by dotted lines in Figure 5. For numerical integration the fully implicit Euler method was used, due its unconditional stability, which made it possible to closely approach the time of collapse (if any). To assure the objectivity of numerical results three different initial values were selected: $(A_0 = 3R_0, B_0 = 3R_0)$, $(A_0 = 3R_0, B_0 = R_0)$, $(A_0 = R_0, B_0 = 3R_0)$. Each time evolution in Figure 5 is supplied with arrows indicating the sign of the corresponding undercooling. For example, the region with $\Delta T_A > 0$ is located above the thick solid curve. We see that in all three cases the cross section approaches the equilibrium state given in Table 1. One temporal evolution $(A_0 = 3R_0, B_0 = 3R_0)$ is also illustrated on the right-hand side of Figure 5.

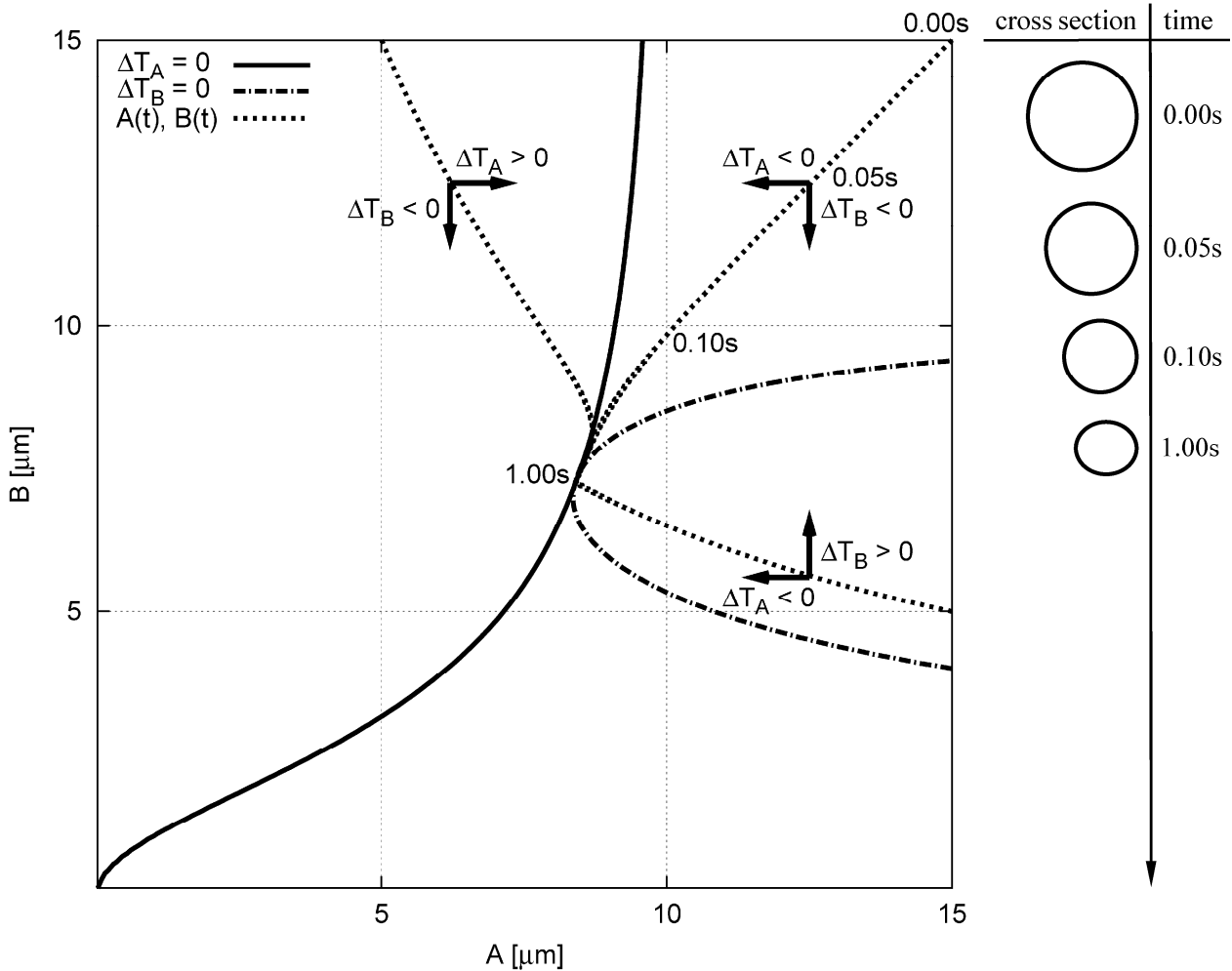

**Figure 5** Results for the critical bending moment $M_x = 0.017\,\text{N}\mu\text{m}$: implicit plots showing lines of zero undercoolings and domains, where they have different signs; time evolution of the neck cross section for different initial values $(3R_0, 3R_0)$, $(3R_0, R_0)$, $(R_0, 3R_0)$ are shown by dotted lines; the case $(3R_0, 3R_0)$ is further illustrated on the right-hand side of the figure. No catastrophic elastic remelting is observed.

## 5.2 Catastrophic loading level

Any further increase of the bending moment above the critical value $M_x^{Cr}$, will prevent the zero undercooling curves from intersection (see Figure 6). Thus the equations system (15) with vanishing left-hand sides will posses no equilibrium states (negative and imaginary solutions were excluded). Physically this means that the stabilizing effect of the negative outer curvature is <u>not</u> sufficient to compensate the growth of elastic undercooling. Independent on the initial value, the time evolution of the dendrite cross section is characterized by catastrophic melting rates according to our qualitative estimation given in Equation (17). The corresponding evolution of the dendrite neck cross section is again shown on the right-hand side. Starting from the circular cross section, the semi-axis B remelts until it vanishes meaning the detachment of the dendrite arm. The time integration was stopped as soon as the inner iterations of Euler scheme for some time increment diverged or converged to a non-physical value for some time increment.

To summarize the results, three time histories are shown also in Figure 7 as explicit plots. In all cases the initial value was the same $(A_0 = 3R_0, B_0 = 3R_0)$, but different loading levels were chosen: catastrophic, critical and stable. The stable loading was taken slightly smaller, than the critical one. In absence of catastrophic elastic remelting both semi-axes of the ellipse converge to some final values close to the $8\mu m$. The values of the B-axis, where the elastic undercooling term is not zero, are always smaller as it can be expected. Under catastrophic loading an increasingly rapid reduction of the B-semi-axis in a rather short time interval of about $t_C \sim 0.3\,\text{sec}$ is clearly observed.

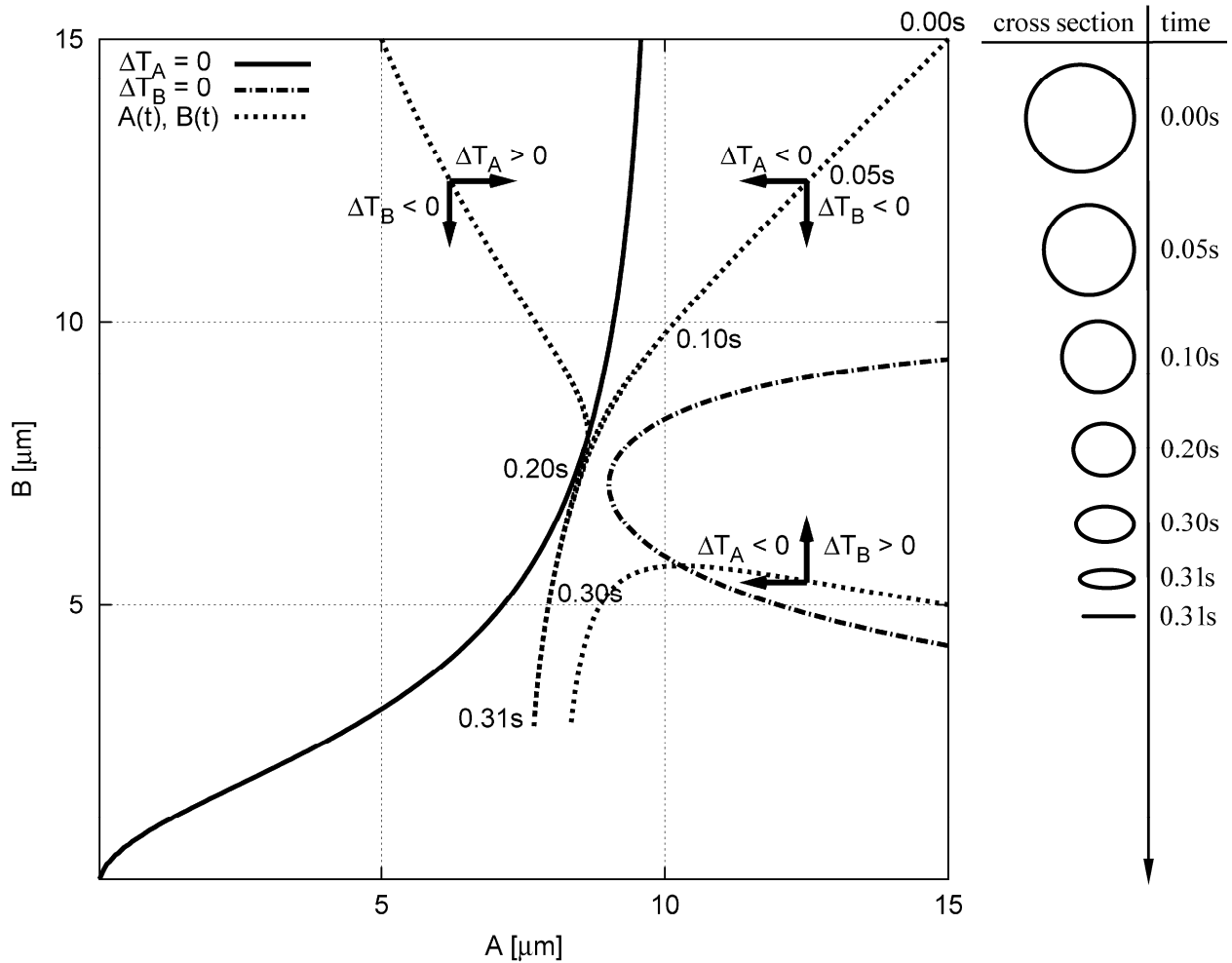

**Figure 6** Results for the catastrophic bending moment $M_x = 0.019\,\text{N}\mu\text{m}$ : implicit plots showing lines of zero undercoolings and domains, where they have different signs; time evolution of the neck cross section for different initial values $(3R_0, 3R_0)$, $(3R_0, R_0)$, $(R_0, 3R_0)$ are shown by dotted lines; the case $(3R_0, 3R_0)$ is further illustrated on the right-hand side of the figure. The catastrophic elastic remelting is observed in all three cases.

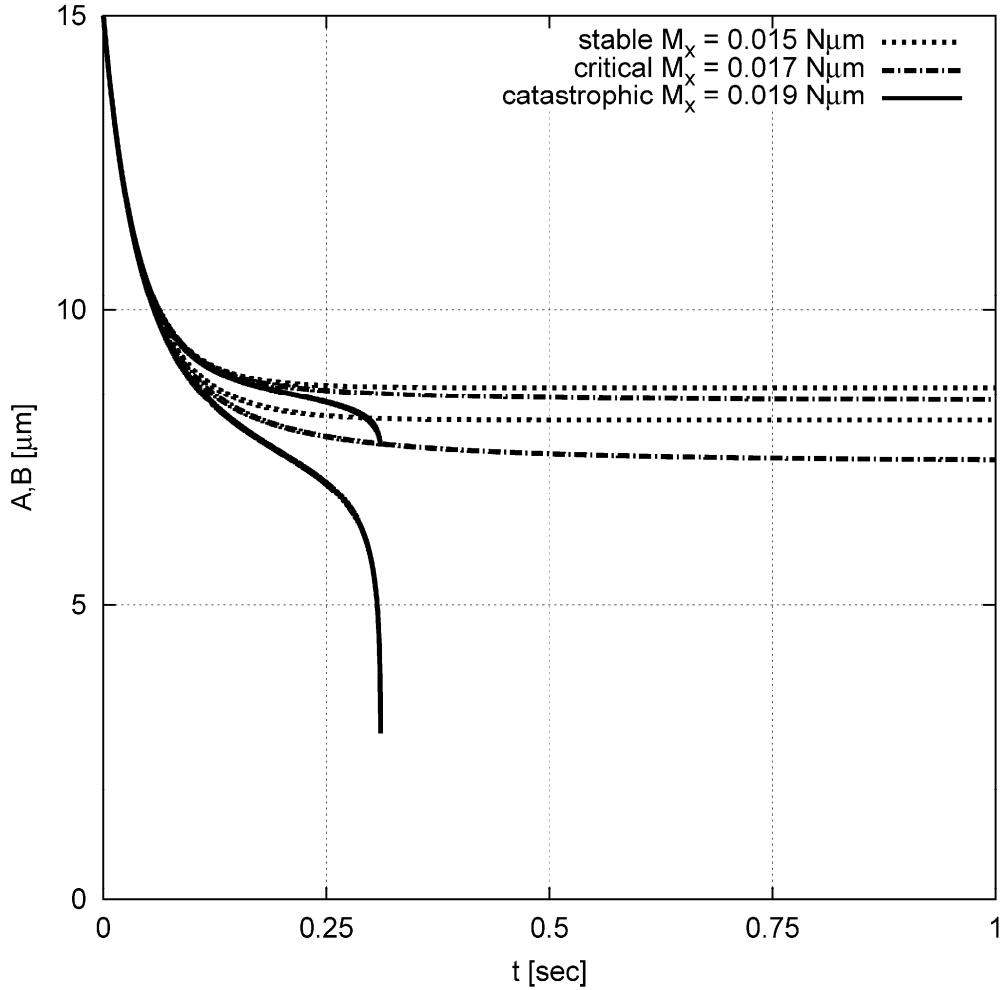

**Figure 7** Time evolution of the dendrite neck cross section for all three loading levels starting from the same initial value $(3R_0, 3R_0)$.

## 6    DISCUSSION

To complete the presentation of our fragmentation mechanism it remains to calculate the maximal stress corresponding to the critical loading found in Section 5.1. Substituting the results from Table 1 in Equation (7) leads to the following value:

$$\sigma^{MAX} = \sigma_{zz}(0, B) = \frac{4M_0}{\pi AB^2} = \frac{4 \cdot 0.017 [\text{N}\mu\text{m}]}{\pi \cdot 8.423 [\mu\text{m}] \cdot (7.309)^2 [\mu\text{m}]^2} \approx 50 [\text{MPa}] \qquad (19)$$

This value is still much higher $(\sim 10^3)$ than the elastic stresses acting in the dendrite under typical growth conditions, where they were estimated to have a value of $\sim 0.06$ MPa (see Figure 3 in [7]). We have to remember however that our analysis was based on an oversimplified geometric model of the dendrite arm (Figure 2). In particular, a critical component of the model was the dependence of the outer curvature $K^{II}_{\{A,B\}}$ on the current value of the corresponding semi-axis. If, for example, we use $K^{II}_0 = 2.0/R_0$, then there will be no equilibrium states at all, implying an immediate fragmentation of the dendrite arm. For a slightly higher value $K^{II}_0 = 2.1/R_0$ the critical bending moment is 10 times <u>lower</u>, than those found in Section 5.1. It seems to us that the question concerning the type of functional dependence for the outer curvature, can only be answered by computational models based on Finite Element or Finite

Volume methods, which allow the elastic and geometric terms in Equation (2) to be calculated for arbitrary shapes and growth conditions of dendrites. This we plan to accomplish in the nearest future.

## 7  SUMMARY AND CONCLUSIONS

In contrast to foregoing works we have shown that mechanical loading can indeed play a substantial role in dendrite fragmentation. By exceeding of some critical level it leads to a *catastrophic elastic remelting (CER)* occurring within a short period of time ($\sim 0.3\,\mathrm{sec}$, see Figure 7). In experiments this can look as a result of mechanical damage (it takes place short after the mechanical loading was applied). The physics behind the CER is however completely different – it is of *constitutional remelting* type. As soon as there is a buoyancy force, the stress state in the solid phase will be altered together with $C_L^*$ along the S-L interface. This can lead to a positive solute gradient resulting in remelting (Figure 4b). But on the other hand, the buoyancy force will cause a melt convection, which can lead to a positive solute gradient as well. This makes us to assume that the CER operates also in the experiments published so far. For its experimental verification it will be therefore important to clearly separate it from the convection induced remelting. This can be achieved if, for example, the solidification velocity is considerably larger than the characteristic velocity of melt flow. Under these conditions dendrites will have enough time for their dead weight to exceed the critical loading level, but at the same time the natural convection will not be able to considerably alter the solute distribution in the melt.

## ACKNOWLEDGEMENTS

We thank Sven Eckert and Stefan Boden for stimulating discussions. This work was financially supported by Deutsche Forschungsgemeinschaft in the form of the collaborative research center SFB 609 "Electromagnetic Flow Control in Metallurgy, Crystal Growth and Electrochemistry".